\documentclass{PoS}

\usepackage{epsf}
\usepackage{graphicx}

\def\rmsub#1#2{#1_{\mbox{\tiny #2}}}	    
\def\rmsup#1#2{#1^{\mbox{\tiny #2}}}	    

\def\naive{na\"{\i}ve}			    
\def\exp{\mathop{\rm exp}}		    
\def\tr{\mathop{\rm tr}}		    
\def\ln{\mathop{\rm ln}}		    
\def\det{\mathop{\rm det}}		    

\def\rational#1#2{{\mathchoice{\textstyle{#1\over#2}}%
 {\scriptstyle{#1\over#2}}{\scriptscriptstyle{#1/#2}}{#1/#2}}}
\def\half{\rational12}			    
\def\quarter{\rational14}		    
\def\eighth{\rational18}		    
\def\RHMC{RHMC}				    
\def\Ralg{R}				    
\def\M{{\cal M}}			    
\def\D{{\mathcal{D}}}			    
\def\dt{\delta\tau}			    
\def\Sf{\rmsub{S}{F}}			    
\def\mpv{{\rmsub{M}{PV}^\dagger \rmsub{M}{PV}}} 
\def\mf{{\rmsub{M}{PF}^\dagger \rmsub{M}{PF}}} 
\def\mdwf{{\rmsup{\M}{DWF}}}		    
\def\ml{\rmsub{m}{ud}}			    
\def\ms{\rmsub{m}{s}}			    

\makeatletter
\newcommand\figcaption{\def\@captype{figure}\caption}
\newcommand\tabcaption{\def\@captype{table}\caption}
\makeatother

\PoS{PoS(LAT2005)115}

\title{Algorithm Shootout: {\Ralg} versus \RHMC}

\ShortTitle{Algorithm Shootout: {\Ralg} versus \RHMC}

\author{\speaker{M.~A.~Clark}\footnote{RBC and UKQCD collaborations.}\\
 Center for Computational Science, Boston University,\\ 
 3 Cummington Street, Boston, MA 02215, United States of America\\
 E-mail: \email{mikec@bu.edu}}

\author{Ph.~de~Forcrand\\
 Institute for Theoretical Physics, ETH Z\"urich,
 CH-8093 Z\"urich, Switzerland\\
 CERN, Physics Department, TH Unit,
 CH-1211 Gen\`{e}ve 23, Switzerland\\
 E-mail: \email{forcrand@phys.ethz.ch}}

\author{A.~D.~Kennedy\footnotemark[\value{footnote}]\\
 School of Physics, The University of Edinburgh,\\
 Mayfield Road, Edinburgh EH9 3JZ, United Kingdom\\
 E-mail: \email{adk@ph.ed.ac.uk}}

\abstract{We present initial results comparing the {\RHMC} and {\Ralg}
algorithms on large lattices with small quark masses using chiral fermions. We
also present results concerning staggered fermions near the
deconfinement/chiral phase transition. We find that the {\RHMC} algorithm not
only eliminates the step-size error of the {\Ralg} algorithm, but is also
considerably more efficient. We discuss several possibilities for further
improvement to the {\RHMC} algorithm.}

\FullConference{XXIII\({}^{\hbox{\it\small rd}}\) International
		 Symposium on Lattice Field Theory\\
		 25-30 July 2005\\
		 Trinity College, Dublin, Ireland}

\begin{document}

\section{Introduction and Motivation}

\noindent We start with an observable \(\Omega\) measured in the vacuum,
\(\langle\Omega\rangle = \int\D U e^{-S(U)} [\det\M(U)]^\alpha \Omega(U)/Z\),
where the parameter \(\alpha\) defines the number of fermion flavours the
theory represents (\(\alpha = N_f/4\) for staggered and \(N_f/2\) for
Wilson). We choose to include the operator \(\M=M^\dagger M\) as opposed to
just \(M\) (the Dirac operator) to allow us to update the pseudofermion fields
using a heatbath. When \(\alpha\) is non-integer, the conventional HMC
algorithm fails, and this necessitates another algorithm choice if we are to
include the effects due to the strange quark (or even if we want to simulate a
two flavour staggered fermion theory).

The most popular algorithm to date for performing such calculations has been
the {\Ralg} algorithm (\Ralg) \cite{Gottlieb:1987mq}. Here the fermionic
determinant is rewritten in exponential-trace-logarithm form \(\det\M^{\alpha}
= \exp\left(\alpha\tr\ln \M\right)\). When the equations of motion for this
effective action are derived, it is found that the explicit matrix inverse is
required. To circumvent this problem the trace is replaced by a noisy
estimator, this results in a fermionic force term almost identical to that
obtained from HMC where a pseudofermion formulation is used. The key difference
is that in the former the force is merely a noisy estimator of the true fermion
force, whereas in the latter the derived force is exactly the pseudofermion
force. This means that the \(O(\dt)\) error term in the conserved Hamiltonian
no longer cancels automatically, and to force this cancellation an irreversible
and non-area preserving integrator must be used. Thus the algorithm cannot be
made exact through the inclusion of a Monte Carlo acceptance test and any
results generated using {\Ralg} will have finite \(O(\dt^2)\) errors.

\section{Rational Hybrid Monte Carlo}

\noindent The Rational Hybrid Monte Carlo Algorithm (\RHMC) was designed to
overcome the shortcomings of \Ralg, namely the inexact nature of the
calculations (for a proper description of the algorithm see
\cite{Clark:2004cp}, the description given here is sparse on details). The
fermion determinant is rewritten in pseudofermion formulation, and a rational
approximation is used to represent the effective matrix appearing in the
bilinear term, \(\det\M^{\alpha} = \int \D\bar\phi \D\phi e^{-\bar\phi
\M^{-\alpha} \phi} \approx \int \D\bar\phi \D\phi e^{-\bar\phi r^2(\M) \phi}\),
with \( r(x) \approx x^{-\alpha/2} \) \footnote{This seemingly perverse choice
of including \(r^2\) in the action is to allow for heatbath evaluation of the
pseudofermion fields.}. If the error \(\Delta\) on the rational approximation
is made arbitrarily small, e.g., \(\Delta(x) = \left|1 - x^{\alpha/2}
r(x)\right| < 10^{-14}\), then there can be no systematic bias induced from
using a rational approximation and the conventional HMC algorithm can be used,
but with the rational function used as the kernel in the bilinear.

Any rational function can be written as a sum of partial fractions, \(r(x) =
\sum_{k=1}^{n} \alpha_k/(x+\beta_k)\), in this form a multi-shift solver
can be used to evaluate all shifts for approximately the same cost as the
smallest shift (which is near zero). A lower degree approximation can be used
for the MD evolution since any errors are corrected for stochastically when the
accept/reject test is performed. This approximation is generally set to
\(\bar{r} \approx \M^{-\alpha} \approx r^2\) to avoid the double inversion
which would have arisen from using \(r^2\). The resulting pseudofermion force
is written
\begin{equation}
  \Sf' = -\sum_{i=1}^{\bar n} \bar\alpha_i
    \phi^\dagger (\M + \bar\beta_i)^{-1} \M' (\M + \bar\beta_i)^{-1} \phi.
\label{eq:rhmc-force}
\end{equation}
Hence the cost of the algorithm is similar to HMC in that it only requires one
conjugate gradient inversion per MD step.

\section{Finite temperature QCD}

\begin{figure}[htb]
 \begin{minipage}[b]{0.48\textwidth}
 \epsfxsize=1.0\textwidth {\epsfbox{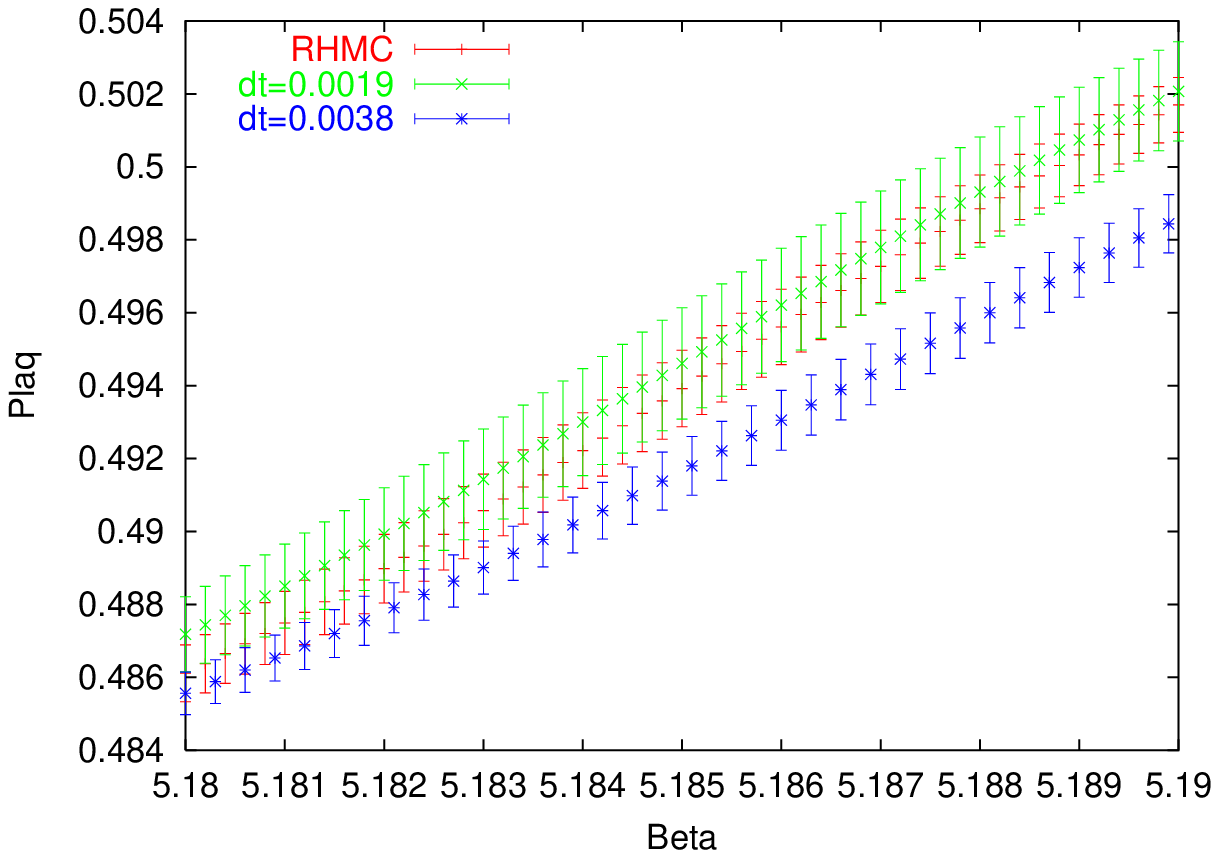}}
 \caption[Figure 1]{The plaquette behaviour near the chiral transition point
                    (parameters given in Table~\ref{tab:finite}).}
 \label{fig:finite-plaq}
 \end{minipage}
 \hfill
 \begin{minipage}[b]{0.48\textwidth}
 \epsfxsize=1.0\textwidth {\epsfbox{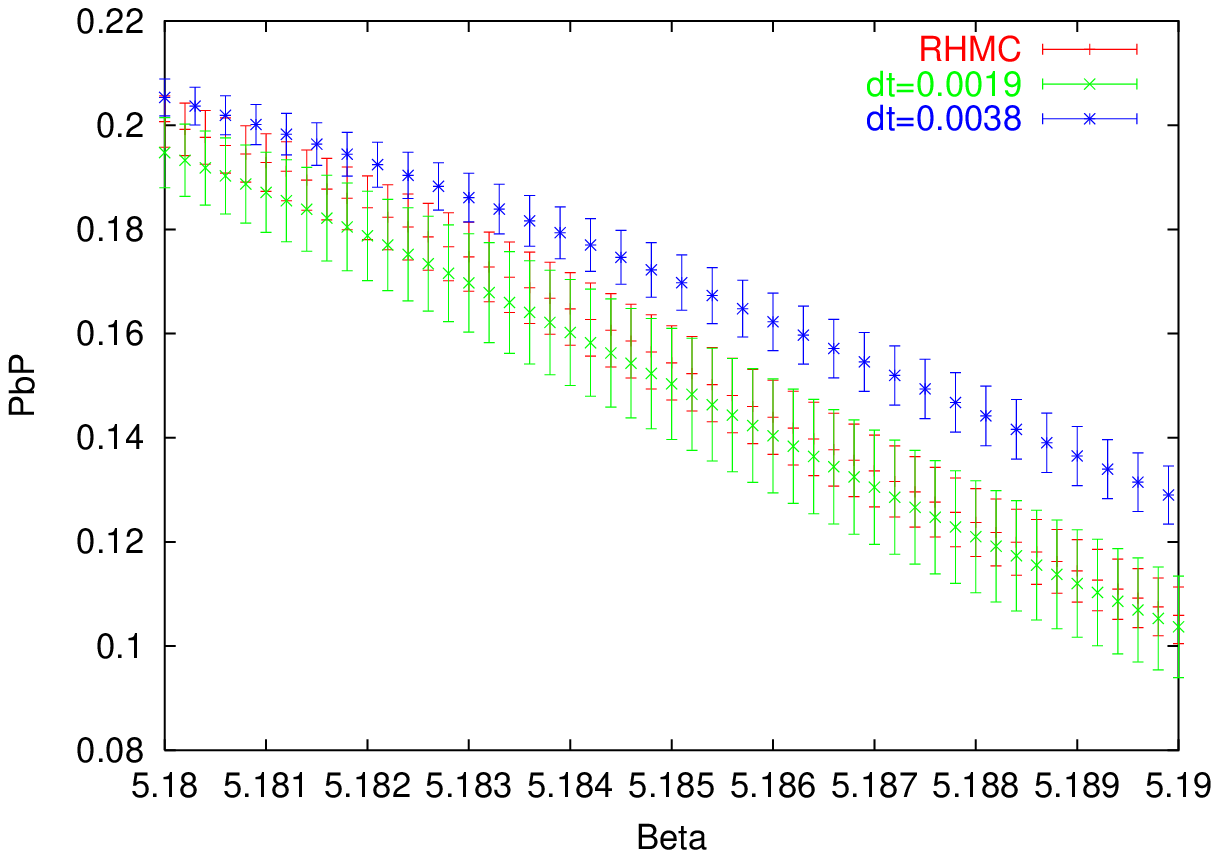}}
 \caption[Figure 2]{The \(\bar{\psi}\psi\) behaviour near the chiral transition
                    point (parameters given in Table~\ref{tab:finite}).}
 \label{fig:finite-pbp}
 \end{minipage}
\end{figure}

\noindent The comparison of the algorithms' performance near the chiral
transition point is of particular relevance. In this regime it has been shown
that the exact location of the transition point \(\beta_c\) can be strongly
affected by finite step-size errors \cite{Kogut:2005qg}. The aim of our study
was to see how {\RHMC} behaves in this regime, and to compare the cost against
using \Ralg.  The parameters we used are given in Table~\ref{tab:finite}. In
Figures~\ref{fig:finite-plaq} and~\ref{fig:finite-pbp} plots of the variation
of the plaquette and of \(\bar{\psi}\psi\) are given: only for the smaller
step-size is {\Ralg} consistent with \RHMC.

\begin{table}[htb]
 \begin{minipage}[b]{0.4\textwidth}
 \begin{center}
 \begin{tabular}{|l|l|l|l|}\hline
	Alg & \(\dt\) & \(A \%\) & \(B_4\) \\ \hline
	\Ralg & 0.0019 & - & 1.56(5) \\ \hline 
	\Ralg & 0.0038 & - & 1.73(4) \\ \hline
	\RHMC & 0.055 & \(\approx 84\) & 1.57(2) \\ \hline
 \end{tabular}
 \caption[Table 1]{Binder cumulant of \(\langle\bar\psi\psi\rangle\), \(B_4\),
                   and {\RHMC} acceptance rate \(A\) from the finite
                   temperature study (\(2+1\) flavour \naive\ staggered
                   fermions, Wilson gauge action, \(V=8^3\times4\),
                   \(\ml=0.0076\), \(\ms=0.25\), \(\tau = 1.0\))}
 \label{tab:finite}
 \end{center}
 \end{minipage}
 \hfill
 \begin{minipage}[b]{0.58\textwidth}
 \begin{center}
 \begin{tabular}{|l|l|l|l|l|l|} \hline
	Alg & \(\ml\) & \(\ms\) & \(\dt\) & \(A\)\% & \(P\)\\ \hline
	\Ralg & 0.04 & 0.04 & 0.01 & - & 0.60812(2)\\ \hline
	\Ralg & 0.02 & 0.04 & 0.01 & - & 0.60829(1)\\ \hline
	\Ralg & 0.02 & 0.04 & 0.005 & - & 0.60817(3)\\ \hline
	\RHMC & 0.04 & 0.04 & 0.02 & 65.5 & 0.60779(1)\\ \hline
	\RHMC & 0.02 & 0.04 & 0.0185 & 69.3 & 0.60809(1)\\ \hline
 \end{tabular}
 \end{center}
 \caption[Table 2]{The different masses at which the domain wall results were
                    gathered, together with the step-sizes \(\dt\), acceptance
                    rates \(A\) and plaquettes \(P\) (\(V=16^3 \times32 \times
                    8\), DBW2 gauge action, \(\beta = 0.72\)).}
 \label{tab:dwf}
 \end{minipage}
\end{table}

On each configuration the Binder cumulant was measured. The cumulant is
important because it probes the strength of the phase transition.  Only when
{\Ralg} is using an integration step-size of \(\dt\le\ml/4\) is the step-size
sufficiently small that the finite step-size errors are negligible (see
Table~\ref{tab:finite}). The correct value is obtained with {\RHMC} using a
step-size $\approx29$ times larger: this represents a considerable improvement
in efficiency.

\section{2+1 Domain Wall Fermions} \label{sec:dwf}

\noindent Again a comparison of {\Ralg} and {\RHMC} was performed, but now
using domain wall fermions with a more realistic volume (the parameters from
this study are presented in Table~\ref{tab:dwf}).  Applying {\RHMC} to the case
of domain wall fermions is not as trivial as for staggered fermions. The one
flavour domain wall fermion determinant is given by \(\left.\det\sqrt\mf\right/
\det\sqrt\mpv = \det\sqrt\mdwf\). The additional Pauli-Villars bosonic
determinant is required to cancel the heavy modes appearing in the bulk of the
fifth dimension.  Unfortunately, we cannot use \(r(\mdwf)\) as this would lead
to a nested inversion in the solver. Therefore, the action is written \(\Sf =
\bar{\phi} \left(\mpv \right)^\half \phi + \bar{\chi} \left(\mf\right)^{-\half}
\chi\), and now each matrix kernel can be written as a rational approximation.
Thus we require 2 fermion fields to simulate a single flavour contribution. The
\naive\ additional cost of this formulation is small (\(\rmsub{m}{PV}\gg\ml\)),
but is inherently more noisy because the heavy mode cancellation is only done
stochastically. This results in larger forces, and a smaller step-size will be
required than if the cancellation was exact \cite{Aoki:2004ht} (the resolution
to this problem shall be presented in \S\ref{sec:exact}). The {\Ralg} algorithm
also uses stochastic cancellation, the bosonic Pauli-Villars field is included
through the use of negative flavour number.

\begin{figure}[htb]
 \begin{minipage}[b]{0.48\textwidth}
 \epsfxsize=0.85\textwidth {\epsfbox{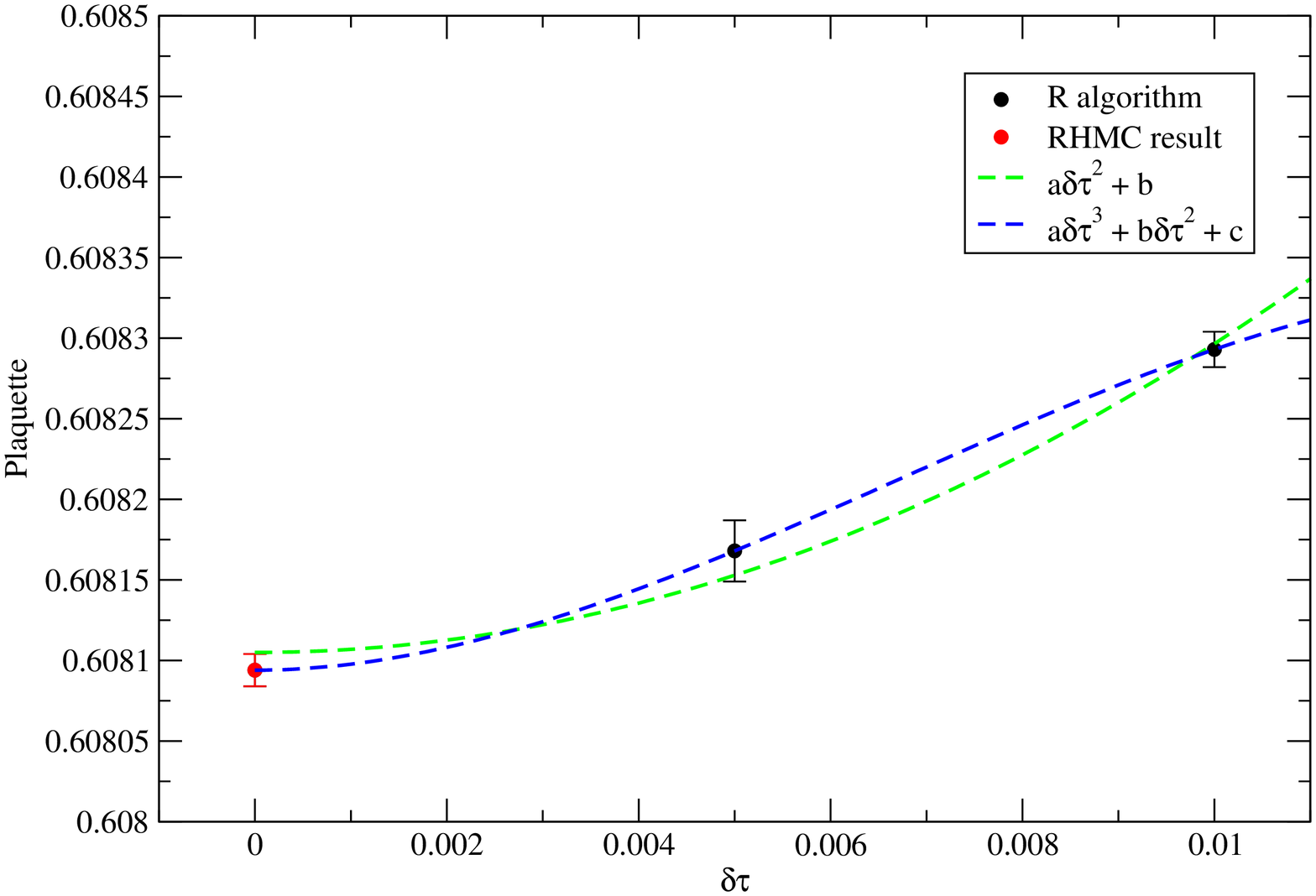}}
 \caption[Figure 3]{The step-size variation of the plaquette (\(\ml=0.02\),
                    additional parameters given in Table~\ref{tab:dwf}). }
 \label{fig:dwf-plaq}
 \end{minipage}
 \hfill
 \begin{minipage}[b]{0.48\textwidth}
 \epsfxsize=0.85\textwidth {\epsfbox{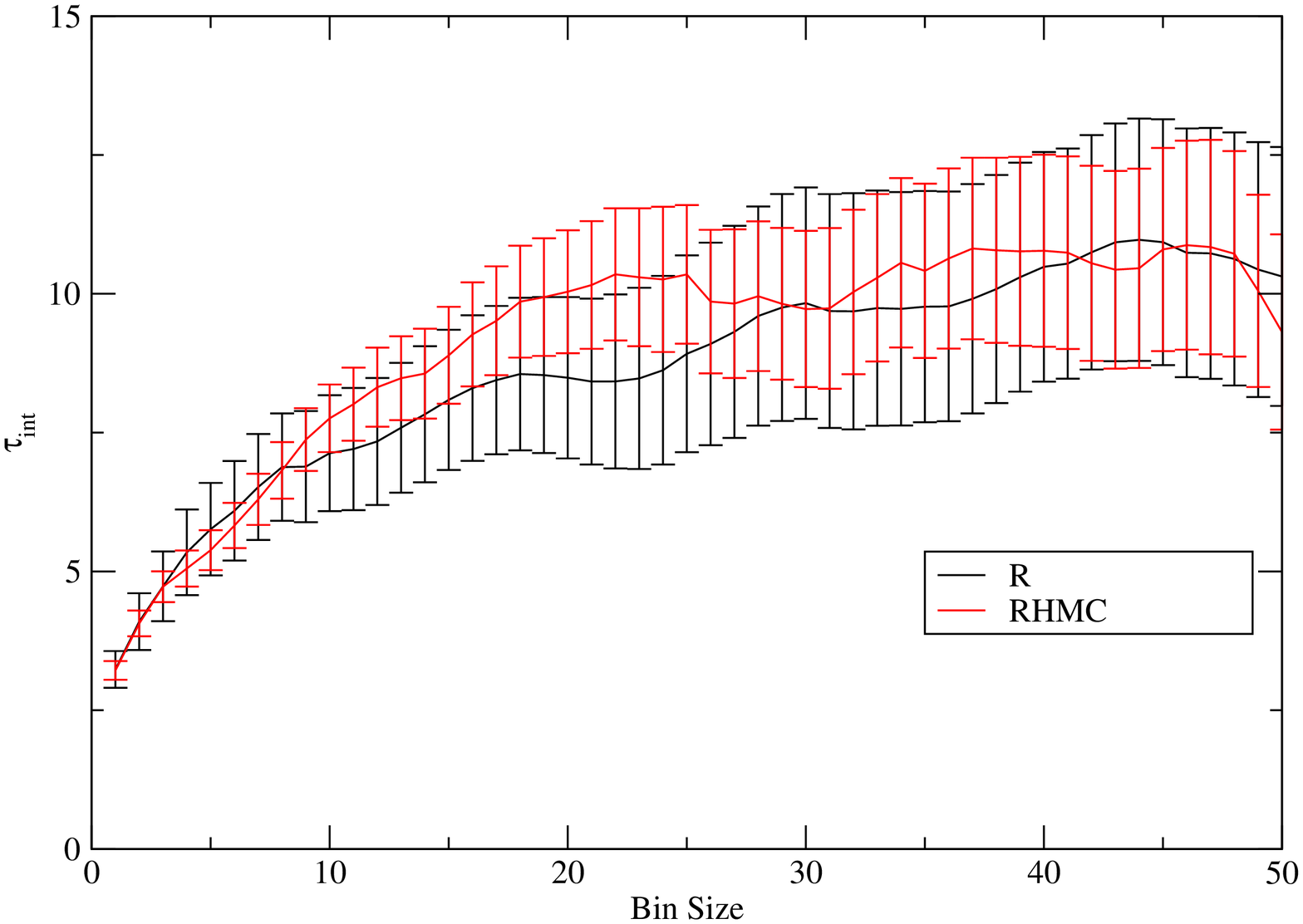}}
 \caption[Figure 4]{The integrated autocorrelation time of the \(13^{\mbox{\rm
                    \tiny th}}\) time-slice of the pion propagator
                    (\(\ml=0.04\), parameters given in Table~\ref{tab:dwf}). }
 \label{fig:dwf-pion}
 \end{minipage}
\end{figure}

The step-size dependence of the plaquette is shown in
Figure~\ref{fig:dwf-plaq}, from this extrapolation it is clear that to obtain a
consistent result between the algorithms requires that {\Ralg} use an
integration step-size at least 10 times smaller than \RHMC.

No algorithm comparison would be complete without a comparison of
autocorrelations. In this study, both the plaquette and pion integrated
autocorrelation times were measured, and it was found that there was very
little to distinguish the two algorithms (see, e.g.,
Figure~\ref{fig:dwf-pion}).

\section{Improving \RHMC}

\subsection{Integration Scheme}

\begin{figure}[htb]
  \begin{minipage}[b]{0.48\textwidth}
  \begin{center}
  \begin{tabular}{|l|l|l|} \hline
  Integrator & \(\dt\) & \(A \%\) \\ \hline 
  Leapfrog & 0.02 & 63.6 \\ \hline
  Omelyan & 0.04 & 88.8 \\ \hline
  \end{tabular}
  \end{center}
  \tabcaption{A comparison of the leapfrog and Omelyan integrators, using
    domain wall fermions (\(V=16^3\times32\times8\), Iwasaki gauge action,
    \(\beta = 2.13\), \(\ml=\ms/2\)).}
  \label{tab:integrator}
  \end{minipage}
  \hfill
  \begin{minipage}[b]{0.48\textwidth}
   \epsfxsize=0.85\textwidth \centerline{{\epsfbox{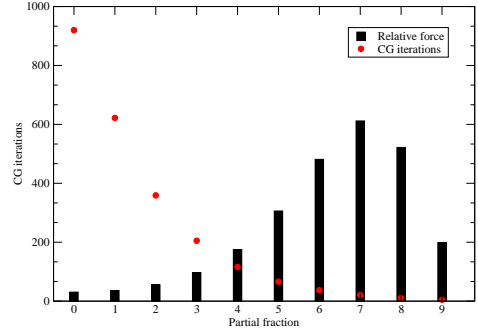}}}
   \caption[Figure 5]{DWF Force magnitude (\(\beta=2.13\), \(\ml/\ms=0.25\),
                      \(V=24^3 \times 64 \times 16\))}
   \label{fig:dwf-force}
  \end{minipage}%
\end{figure}

\noindent It has recently been shown \cite{Takaishi:2005tz}, that the optimal
second order symmetric symplectic integrator is not the leapfrog integrator,
rather it is that given by Omelyan {\it et al} \cite{Omelyan:2003}, which is
given by
\begin{displaymath}
  \rmsub{\hat{U}}{QPQPQ}(\dt) = e^{\lambda\dt Q} \; e^{\dt P/2}
    \; e^{(1-2\lambda)\dt Q} \; e^{\dt P/2} \; e^{\lambda\dt Q},
\end{displaymath}
where \(Q\) and \(P\) represent the coordinate and momenta update operators
respectively and \(\lambda\) is a tuneable parameter whose optimal value is
found to be \(\lambda \approx 0.1931\). The Omelyan integrator is approximately
double the cost of leapfrog, but theoretically should lead to a 3 fold
improvement in conservation of the Hamiltonian: a net 50\% gain. This
integration scheme can of course be used with \RHMC, and leads to a near 40\%
improvement in acceptance rate (see Table~\ref{tab:integrator}).

\subsection{Exact Heavy Mode Cancellation} \label{sec:exact}

\noindent In the results presented in \S\ref{sec:dwf}, the {\RHMC} domain wall
calculations were performed using only a stochastic cancellation of the heavy
modes, and it was noted that this leads to larger fermion forces, hence a more
costly algorithm. Since that study was conducted, a resolution to this problem
has been found. The one flavour DWF determinant can be rewritten as
\begin{displaymath}
  \sqrt{\frac{\det\mf}{\det\mpv}}
    = \det \left[(\mpv)^{-\eighth} (\mf)^{\quarter} (\mpv)^{-\eighth}\right]^2,
\end{displaymath}
and the resulting pseudofermion action is given by
\begin{displaymath} 
  \Sf = \bar{\phi}
      \left[(\mpv)^\quarter (\mf)^{-\half} (\mpv)^\quarter\right]^2 \phi 
    = \bar{\phi}\, \left[r_1(\mpv)\, r_2(\mf)\, r_1(\mpv)\right]^2\, \phi;
\end{displaymath}
where \(r_1(x)\approx x^\quarter\) and \(r_2(x)\approx x^{-\half}\). The kernel
that appears in the bilinear term is now in a form that allows heatbath
evaluation (it is the square of a real positive operator) and a multi-shift
solver can be used to evaluate each rational function that appears in the
action. At each step of the MD trajectory three inversions are required
compared to two for the stochastic formulation presented in
\S\ref{sec:dwf}. Since two of these inversions are using the Pauli-Villars
matrix the cost of the extra inversion is negligible. This formulation should
allow for large increases in the integration step-size. This shall be the
subject of further study.

\subsection{Fermion force tuning}

\noindent The {\RHMC} fermionic force given in equation~(\ref{eq:rhmc-force})
is just a sum of HMC-like force terms, each with different
magnitude. Figure~\ref{fig:dwf-force} is a plot showing the magnitude of the
force associated with each shift. Also included is the number of conjugate
gradient iterations required by the solver for each shift. The key point is
that the most expensive shifts are also those which contribute least to the
total fermion force. Since the acceptance rate is determined by the quantity
\(F\dt\), this suggests that the small shifts can use a larger integration
step-size than the large shifts. Hence, we could split the partial fractions
into multiple timescales and use a multi-timescale integration scheme. An
alternative strategy consists of relaxing the convergence of the multi-shift
solver for the smaller shifts, but maintaining tight convergence for the larger
shifts. An initial study into how much improvement can be gained from these
strategies is currently being undertaken, it would appear that at least a
factor of two can be gained, subject to further tuning.

\section{Conclusions and outlook}

\noindent In this work we have presented the exact {\RHMC} algorithm and
compared it to the inexact {\Ralg} algorithm in two regimes, namely near the
chiral transition point of QCD using small volumes, and at low temperature
using more realistic volumes. In both regimes, it was found that {\Ralg} must
be run using a much smaller step-size than {\RHMC} to achieve consistency
between the two algorithms. Extrapolation of the {\Ralg} results to zero
step-size is also significantly more expensive than just using \RHMC. The
conclusion therefore, is that there is no need for further use of the {\Ralg}
algorithm.

Various improvements to the standard {\RHMC} formulation were presented which
lead to further performance improvements. It would be interesting to compare
this improved {\RHMC} to other exact algorithms, indeed this shall be the
subject of future study.

\section{Acknowledgements}
\noindent We thank Chris Maynard and Rob Tweedie for help generating the
datasets used in this work. We thank Peter Boyle, Dong Chen, Norman Christ,
Saul Cohen, Calin Cristian, Zhihua Dong, Alan Gara, Andrew Jackson, B\'alint
Jo\'o, Chulwoo Jung, Richard Kenway, Changhoan Kim, Ludmila Levkova, Xiaodong
Liao, Guofeng Liu, Robert Mawhinney, Shigemi Ohta, Konstantin Petrov, Tilo
Wettig, and Azusa Yamaguchi for developing with us the QCDOC machine and its
software. This development and the resulting computer equipment used in this
calculation were funded by the U.S. DOE grant DE-FG02-92ER40699, PPARC JIF
grant PPA/J/S/1998/00756 and by RIKEN. This work was supported by PPARC grant
PPA/G/O/2002/00465. Ph.~de~F. thanks the Minnesota Supercomputer Institute for
computer resources.

\end{document}